\documentstyle[12pt,epsf]{article}
\begin{document}
\setlength{\baselineskip}{0.33 in}
\catcode`@=11
\long\def\@caption#1[#2]#3{\par\addcontentsline{\csname
  ext@#1\endcsname}{#1}{\protect\numberline{\csname
  the#1\endcsname}{\ignorespaces #2}}\begingroup
    \small
    \@parboxrestore
    \@makecaption{\csname fnum@#1\endcsname}{\ignorespaces #3}\par
  \endgroup}
\catcode`@=12
\newcommand{\newc}{\newcommand}
\newc{\gsim}{\lower.7ex\hbox{$\;\stackrel{\textstyle>}{\sim}\;$}}
\newc{\lsim}{\lower.7ex\hbox{$\;\stackrel{\textstyle<}{\sim}\;$}}
\def\NPB#1#2#3{Nucl. Phys. {\bf B#1} #3 (19#2)}
\def\PLB#1#2#3{Phys. Lett. {\bf B#1} #3 (19#2)}
\def\PRD#1#2#3{Phys. Rev. {\bf D#1} #3 (19#2)}
\def\PRL#1#2#3{Phys. Rev. Lett. {\bf#1} #3 (19#2)}
\def\PRT#1#2#3{Phys. Rep. {\bf#1} #3 (19#2)}
\def\MODA#1#2#3{Mod. Phys. Lett. {\bf A#1} #3 (19#2) }
\def\ZPC#1#2#3{Zeit. f\"ur Physik {\bf C#1} #3 (19#2) }
\def\bdm{\begin{equation}}
\def\edm{\end{equation}}
\def\bea{\begin{eqnarray}}
\def\eea{\end{eqnarray}}
\def\vef{V_{\rm{eff}}}
\def\lef{\lambda_{\rm{eff}}}
\def\msb{\overline{MS}}
\def\rxiu{R_{\xi,u}}
\def\rxi{R_{\xi}}
\def\pole{\left[{{-\Delta_{\epsilon}} \over {{\left( 4 \pi \right)}^2}} \right]}

\def\eea{\end{eqnarray}}
\def\sl#1{\setbox0=\hbox{$#1$}           
   \dimen0=\wd0                                 
   \setbox1=\hbox{/} \dimen1=\wd1               
   \ifdim\dimen0>\dimen1                        
      \rlap{\hbox to \dimen0{\hfil/\hfil}}      
      #1                                        
   \else                                        
      \rlap{\hbox to \dimen1{\hfil$#1$\hfil}}   
      /                                         
   \fi}                                         %
%
\vsize 8.7in
\def\singlespace{\baselineskip 11.38 pt}
\def\halfagainspace{\baselineskip 17.07 pt}
\def\doublespace{\baselineskip 22.76 pt}
\def\medspace{\baselineskip 17.07 pt}
\font\headings=cmbx10 scaled 1200
\font\title=cmbx10 scaled 1200
\halfagainspace

\begin{titlepage}
\begin{flushright}
{\large
hep-ph/9704425\\
May 12, 1997 \\
}
\end{flushright}
\vskip 2cm
\begin{center}
{\Large {\bf Gauge Invariant Effective Potentials and Higgs Mass Bounds} }
\vskip 1cm
{\Large A. Duncan\footnote{E-mail: {\tt tony@dectony.phyast.pitt.edu}},
Will Loinaz\footnote{E-mail: {\tt loinaz+@pitt.edu}},
R.S. Willey\footnote{E-mail: {\tt willey@vms.cis.pitt.edu}}\\}
\vskip 2pt
{\large\it Department of Physics and Astronomy\\
 University of Pittsburgh, Pittsburgh, PA 15260, USA}\\
\end{center}

\vspace*{.3in}
\begin{abstract}
  The problem of defining a gauge invariant effective potential with a strict
 energetic interpretation is examined in the context of spontaneously broken
 gauge theories. It is shown that such a potential can be defined in terms
 of a composite gauge invariant order parameter in physical gauges. This
 effective potential is computed through one loop order in a model with scalars
 and fermions coupled to an abelian gauge theory, which serves as a simple
 model of the situation in electroweak theory, where vacuum stability arguments
 based on the scalar effective potential have been used to place lower
 bounds on the Higgs mass.
\end{abstract}
\end{titlepage}
\setcounter{footnote}{0}
\setcounter{page}{2}
\setcounter{section}{0}
\setcounter{subsection}{0}
\setcounter{subsubsection}{0}
\setcounter{equation}{0}

\newpage
\section{Introduction}
    Effective potential calculations in electroweak theory have been the subject of sustained interest as a consequence of the observation that 
 phenomenologically interesting lower
  bounds on  the Standard Model Higgs mass can be obtained from vacuum stability considerations
 (see \cite{SherRev} for a comprehensive review of work through 1989, \cite{QuirosRev} for a  review of more recent work).  
 Although  the gauge interactions do not play a qualitatively 
 significant role in these estimates, it is important to include them if  one wishes to
 obtain as precise a bound as possible. Unfortunately, the conventional effective
 potential is an 
 inescapably gauge-variant object, as the scalar field must transform nontrivially under the
 gauge group in order to break the symmetry spontaneously in the first place. 
 While general theorems \cite{Nielsen} ensure that the {\em value}
  of the effective potential at local extrema is a gauge-invariant quantity, 
  the location of the extrema and the behavior of the potential between extrema 
 can vary widely from one gauge to another.  This makes it difficult to reach 
 unambiguous conclusions on the basis of vacuum stability arguments \cite{LW}.

   There is a long history of attempts to formulate spontaneous symmetry breaking (SSB) in
 terms of gauge-invariant order parameters \cite{BanksRaby, Fischler, DolanJackiw}. In the case of
 dynamical symmetry breaking, where the spontaneous breakdown arises from the
 appearance of a vacuum expectation value for a gauge-invariant operator, such order
 parameters have necessarily been taken to be composite operators \cite{GrossNeveu, CJT, BFH}.
  Thus one is led to  the introduction of an effective action for composite gauge-invariant fields,
 defined in the usual way as the Legendre transform of the connected generating 
 functional for  n-point insertions of the composite operator \cite{UKF}. Such approaches can
 further be divided into those based on a local composite operator, and those where the
 order parameter involves a bilocal operator, i.e. a product of fields at separated space-time
 points. In the former case it is well-known \cite{BanksRaby, YH} that the additional subtractions needed to
 render  the n-point functions of the local composite operator finite vitiate the energetic
 interpretation of the effective potential (obtained as the translationally invariant limit of the
 effective action). This is obviously fatal for any attempt to study vacuum stability using
 such potentials.  Other authors have employed bilocal operators \cite{BanksRaby} with gauge-invariantizing
  factors, but in non-physical covariant gauges where a rigorous energy interpretation is again 
 lacking.

   In this paper we present a complete calculation to the one-loop level of a composite
 effective potential  based on a gauge-invariant order parameter in a physical gauge (Coulomb)
 where the energetic interpretation of the potential is preserved. The bilocal operators
 used involve a smearing function which is largely arbitrary.  The smearing 
 dependence of the results can however be understood in a completely physical
 way, and is found to be numerically insignificant  in the regime of interest. 
The formalism needed
 is illustrated first in a simple scalar model (Section 2), and then extended to the
 Higgs-abelian model coupled to fermions (Section 3). The effective potential computed
 in Section 3 is renormalized and the final  explicit finite result for
 arbitrary smearing functions given in Section 4.  Section 5 contains an explicit analytic
 evaluation for the special case of a smearing function with a sharp cutoff in momentum space.
 In Section 6 we apply the composite operator formalism to the issue of electroweak 
 vacuum stability bounds. Renormalization group improvement of the composite effective 
 potential is also discussed here. Some explicit numerical results are presented in 
 Section 7. The issue of extension of the formalism to nonabelian gauge groups leads
 naturally to the subject of composite effective potentials in axial gauge. The choice
 of an appropriate bilocal operator in this case involves a number of subtleties at both 
 the perturbative and nonperturbative level. Some of these issues are discussed in
 Section 8, although a full computation at the one loop level is deferred to a future
 publication \cite{WL}. 

\section{One-loop effective potential for a composite operator:  a simple example}

 In order to establish notation, and to remind the reader of some features of the loop
 expansion of effective potentials which arise when the order parameter is a
 bilocal composite field, we first consider a simple model of a single real scalar field
 with Lagrangian
\begin{equation}
\label{eq:lagrang1}
 {\cal L}=\frac{1}{2}\partial_{\mu}\phi\partial^{\mu}\phi+\frac{1}{2}m_{0}^{2} \phi^2-P(\phi)
\end{equation}
   Conventionally one studies symmetry breaking in this model by constructing an
 effective potential defined \cite{RJ} as the Legendre transform of $W(j)$, the generating
 functional of connected graphs:
\begin{equation}
\label{eq:W1}
 W(j)\equiv -i\log{Z(j)} = -i\log{\int{\cal D}\phi e^{\frac{i}{\hbar}\int ({\cal L}-j\phi)d^{4}x}}
\end{equation}
  The corresponding Legendre transform, $\Gamma(\phi)$ (where now $\phi$ represents a classical field)
 has a straightforward interpretation in terms of one particle irreducible (1PI) graphs and is easily evaluated
 graphically at one-loop order. If we wish instead to add a source for a composite bilocal operator, 
 the graphical interpretation is more complicated, and it turns out to be easier to construct the one-loop
 potential by a direct semi-classical expansion.  The calculation is also readily performed  for a composite 
 operator in which the field point splitting is smeared {\em spatially} (in all directions)
 with a function $K(\vec{r})$, which
 is essentially arbitrary, except for (a) being nonsingular at the origin $\vec{r}=0$ (in particular, there
 must be no delta function singularities there), and (b) having a positive Fourier transform $\tilde{K}(\vec{p})$.
 The reason for the first condition is well-known- $n$-point functions of local composite fields will require
 additional subtractions which correspond to nonlinear source terms which ruin \cite{Erice} the energy interpretation
 of the effective potential. The positivity requirement on $\tilde{K}(\vec{p})$ ensures that if we define our
 composite field as
\begin{equation}
\label{eq:compdef1}
  \chi(x)\equiv \int \phi(x+\vec{r}/2)K(\vec{r})\phi(x-\vec{r}/2)d^{3}\vec{r}
\end{equation}
 then expectation values of $\chi$ are finite (once the field $\phi$ is renormalized) and {\em positive}. The latter
 property is easily seen by inserting a complete set of states between the two $\phi$ fields in $<0|\chi(x)|0>$
 and using translational invariance. This is in distinction to the case of a local composite field, where
 the subtractions needed to define the composite operator would ruin the positivity, as well as
 introducing the need for counterterms nonlinear in the source which ruin the energy interpretation
 of the effective potential. Here we are
 assured that the appropriate domain for the effective potential $V(\chi)$ is $\chi>0$. Finally, it will be convenient to normalize the smearing function $K(\vec{r})$
 by $\int d\vec{r}K(\vec{r})=1$. This ensures that at the classical level, $\chi$ reduces simply to $\phi^2$ in the
 translationally invariant case $\phi(x)={\rm const}$. In momentum space, this means that  we take $\tilde{K}(\vec{p}=0)=1$.
 A convenient choice is the Gaussian $\tilde{K}(\vec{p})=e^{-\vec{p}^2/(4\rho^2)}$ 
where the parameter $\rho$ is roughly the inverse
 smearing separation of the fields in coordinate space. However, it is not essential that $\tilde{K}$
 be a smooth function- one may also use a sharp cutoff in momentum space, with   $\tilde{K}(\vec{p})=\theta(\rho-|\vec{p}|)$, which has the advantage that the one-loop integrals for the
 composite effective potential can be performed analytically. As we shall see below, the qualitative
 results are similar in both cases when $\rho$ takes a physically sensible value.

  The field $\chi$ is not an order parameter for symmetry breaking in the conventional sense, vanishing
 exactly in the symmetric phase and giving a non-zero expectation value in the broken phase. Evidently,
 $\chi\neq 0$ even in the symmetric phase. However, we shall see below that if the smearing scale
 ${1 \over \rho}$ is taken large compared to the inverse mass gap in the theory, the value of $\chi$ at the minimum
 of the potential does go to zero in the symmetric
 phase, while approaching the square of the conventional vacuum expectation value of $\phi$ in the
 broken phase. Consequently, the scale (if any) at which symmetry breaking occurs can just as well
 be studied by looking for global minima of the potential $V(\chi)$, provided that such a potential is defined
 in a manner consistent with an energy interpretation. This means that composite operators {\em must} be
 point-split (to avoid counterterms nonlinear in the source) and the point-splitting must be spatial only,
 thereby maintaining the Schr\"odinger  picture derivation of the energetics of the system perturbed by a
 source $j\chi$.

     The calculation of the Legendre transform  $\Gamma(\chi)$ of  $W(j)$
\begin{equation}
\label{eq:W2}
 W(j)\equiv -i\log{Z(j)} = -i\log{\int{\cal D}\phi e^{\frac{i}{\hbar}\int ({\cal L}-j\chi)d^{4}x}}
\end{equation}
through one loop is most easily performed by saddle-point techniques. As we are interested in the result 
 only through ${\cal O}(\hbar)$, for the calculation of $W(j)$ we need only keep Gaussian fluctuations around
 the saddle-point $\phi_{0}$ of the source-augmented action:
\begin{equation}
\label{eq:saddle1}
 (\Box +m_{0}^2)\phi_{0}(x)+P^{\prime}(\phi_0)+2\int d\vec{r}j(x-\frac{\vec{r}}{2})K(\vec{r})\phi_0(x-\vec{r})=0
\end{equation}
  Passing to the Legendre transform will require the elimination of the source $j(x)$ in favor of the
 transform variable $\chi$.  Taking the Gaussian smearing $\tilde{K}(\vec{p})=e^{-\vec{p}^2/(4\rho^2)}$ to be
 specific, one may write formally (by Taylor expanding the fields $\phi(x\pm\frac{1}{2}\vec{r})$ in the definition
(\ref{eq:compdef1})) 
\begin{equation}
\label{eq:locexp}
 \chi(x) = \phi(x)^2 -\frac{1}{8\rho^{2}}|\vec{\nabla}\phi(x)|^{2}+...
\end{equation}
  Evidently, as asserted previously, in the translationally invariant limit we recover simply $\chi=\phi^2$. Taking
 $P(\phi)=\lambda_0\phi^4/4$, the saddle-point equation (\ref{eq:saddle1}) then implies
\begin{equation}
\label{eq:source1}
  j_{0}=-\frac{1}{2}(m_{0}^{2}+\lambda_{0}\phi_{0}^{2})=-\frac{1}{2}(m_{0}^{2}+\lambda_0\chi_{0})
\end{equation}
 in the translationally invariant limit. Here, the subscript ``0" on the fields $\phi,\chi$ refers to the leading
 order in the $\hbar$ loop expansion, while the subscripts on $m,\lambda$ remind us that these parameters
 are bare ones, with $m_0=m_R+\hbar\cdot{\rm (1-loop\;\; counterterms)}$.  The leading contribution to the 
 effective potential (defined as the effective action, or Legendre transform of $W(j)$ per unit volume in the
 translationally invariant limit where $\chi(x)={\rm constant}$), is obtained by evaluating the classical action
 at the saddle point $\phi_{0}$ and eliminating $\phi_0$ in favor of $\chi=\phi_{0}^{2}+{\cal O}(\hbar)$. As the saddle-point
 is by definition an extremum, the ${\cal O}(\hbar)$ correction in $\chi$ only affects the result at ${\cal O}(\hbar^2)$ and
 so may be neglected to 1-loop order. Not surprisingly, we recover at order $\hbar^{0}$ the expected tree
 potential, this time written as a function of  the classical composite field $\chi$
\begin{equation}
\label{eq:tree1}
  V_{\rm tree}(\chi) = \frac{1}{2}m_{0}^2\chi+\frac{\lambda_{0}}{4}\chi^2
\end{equation}

  The one-loop contribution is obtained by integrating out the quadratic fluctuations around the saddle-point
 field $\phi_0$. Unlike the usual case where the source is coupled to the elementary scalar, so that the source term
 does not contribute to the quadratic part, here the source term acts as a momentum-dependent mass term at the
 one-loop level.  Defining $P_{\rm tot}(\phi)\equiv P(\phi)+j\chi$ the one-loop integral gives a contribution
 to the effective potential
\begin{equation}
\label{eq:v1loop1}
 V_{\rm 1-loop} = -\frac{i}{2}{\rm Tr}\ln{(\Box+m_{0}^{2}+P_{\rm tot}^{\prime\prime}(\phi)|_{\phi_{0}})}
\end{equation}
   For translationally invariant saddle points, the determinant in (\ref{eq:v1loop1}) reduces to (we are
 still in Minkowski space)
\begin{equation}
\label{eq:1loopeq1}
  {\rm Tr}\ln{(\Box+m_{0}^{2}+P_{\rm tot}^{\prime\prime})}=\int\frac{d^{4}p}{(2\pi)^{4}}
\ln{(p^2-m_{0}^{2}-P^{\prime\prime}(\phi_{0})-2j_{0}\tilde{K}(\vec{p}))}
\end{equation}
 where $j_0$ is to be written in terms of $\chi$ using (\ref{eq:source1}), and we may replace $\chi_0$ by $\chi$
 everywhere in (\ref{eq:v1loop1}) as we are already at ${\cal O}(\hbar)$.  After a Wick rotation to Euclidean space,
 the 1-loop contribution may now be written
\begin{equation}
\label{eq:1-loopeuc}
  V_{\rm 1-loop}= \frac{1}{2}\int\frac{d^{4}p}{(2\pi)^{4}}(\ln{(p^2+m_{0}^{2}(1-\tilde{K}(\vec{p}))+3\lambda_{0}\chi-\lambda_{0}\chi
 \tilde{K}(\vec{p}))}-\ln{(p^{2})})
\end{equation}
  The second term subtracts off a field-independent divergence which is physically insignificant. Although we continue to write the 1-loop contribution in terms of the bare parameters $m_0,\lambda_0$, these may be replaced with renormalized
 parameters in the 1-loop term, which is already of order $\hbar$. It will be convenient to perform the renormalization
 of the theory using the $\overline{\rm MS}$ scheme, as we shall shortly be moving on to gauge theories. In this scheme
 the relation between bare and renormalized quantities is (to order $\hbar$) in $4-\epsilon$ dimensions
\begin{eqnarray}
\label{eq:counterterm}
  \lambda_0 &=& \lambda_{R}+\hbar\frac{9}{16\pi^{2}}\lambda_{R}^{2}\Delta_{\epsilon} \\
  m_{0}^{2} &=& m_{R}^{2}(1+\hbar\frac{3}{16\pi^{2}}\lambda_{R}\Delta_{\epsilon})\\
  \Delta_{\epsilon} &=&\frac{2}{\epsilon}-\gamma+\ln{4\pi}
\end{eqnarray}
where $\gamma$ is the Euler constant and the factor of $\hbar$ makes the 1-loop order at which we are
 working explicit. 

As the smearing
 function only involves the spatial components of the four vector $p$, we first integrate out the energy component 
 explicitly, leaving a 3-$\epsilon$ dimensionally regulated integrand:
\begin{equation}
\label{eq:3dint}
 V_{\rm 1-loop}=\frac{1}{2}\hbar\int\frac{d^{3-\epsilon}p}{(2\pi)^{3-\epsilon}}(\sqrt{\vec{p}^{2}+2\lambda_{R}\chi+
(m_{R}^{2}+\lambda_{R}\chi)(1-\tilde{K}(\vec{p}))}-|\vec{p}|)
\end{equation}
 where we can replace bare by renormalized quantities as we are already at 1-loop order. The divergent parts
 of this expression may be isolated by writing

\begin{eqnarray}
\label{eq:exploop1}
 V_{\rm 1-loop}&=&V_{\rm fin}+V_{\rm sub} \\
 V_{\rm fin}&\equiv&\frac{1}{2}\hbar\int\frac{d^{3}p}{(2\pi)^{3}}
(\sqrt{\vec{p}^{2}+2\lambda_{R}\chi+(m_{R}^{2}+\lambda_{R}\chi)(1-\tilde{K}(\vec{p}) )}-|\vec{p}| \nonumber \\
& -&\frac{1}{2|\vec{p}|}(m_{R}^{2}+3\lambda_{R}\chi)
    +\frac{1}{8}\frac{1}{(\vec{p}^{2}+\eta^{2})^{3/2}}(m_{R}^{2}+3\lambda_{R}\chi)^{2})  \\
V_{\rm sub}&=&-\frac{1}{16}\hbar\int\frac{d^{3-\epsilon}p}{(2\pi)^{3-\epsilon}}\frac{(m_{R}^{2}+3\lambda_{R}\chi)^{2}}
{(p^{2}+\eta^{2})^{3/2}}
\end{eqnarray}
 Here $\eta$ is an infrared cutoff introduced to ensure that $V_{\rm fin}$ remains both infrared and ultraviolet
 finite when $\epsilon$ is taken to zero. The first subtraction term vanishes in dimensional regularization.  The subtraction term is easily computed in $3-\epsilon$ dimensions:
\begin{equation}
\label{eq:subterm1}
 V_{\rm sub}=-\frac{1}{64\pi^{2}}\hbar(m_{R}^{2}+3\lambda_{R}\chi)^{2}(\Delta_{\epsilon}-\ln{\frac{\eta^{2}}{\mu^{2}}})
\end{equation}
with $\mu$ the usual renormalization scale which is introduced by the dimensional continuation. Adding together the tree contribution 
 (\ref{eq:tree1}) (with the bare quantities replaced by the renormalized ones using (\ref{eq:counterterm})) one finds
 the renormalized composite field potential through 1-loop, up to an irrelevant additive constant, to be
\begin{equation}
\label{eq:vrenorm1}
 V_{\rm ren}(\chi)=\frac{1}{2}m_{R}^{2}\chi+\frac{1}{4}\lambda_{R}\chi^{2}+V_{\rm fin}(\chi)+\frac{1}{64\pi^{2}}
(m_{R}^{2}+3\lambda_{R}\chi)^{2}\ln{\frac{\eta^{2}}{\mu^{2}}}
\end{equation}
The logarithmic dependence on $\eta$ in the last term is cancelled by a similar dependence in $V_{\rm fin}$, so
 we can take $\eta$ to zero after the integrals are performed. 

  The local limit for the composite field corresponds to taking $\rho$ in  $\tilde{K}(\vec{p})=e^{-\vec{p}^2/(4\rho^2)}$
 to infinity, i.e. $\tilde{K} \rightarrow 1$. If we do this, a quadratic divergence reappears in $V_{\rm fin}$, as the
 $\chi$-dependence at large momentum no longer matches that of the counterterms needed to renormalize the
 theory. This is a well-known difficulty, much discussed in the literature. Alternately, if we take 
 $\rho$ small, corresponding to separating the field points in the bilocal composite by large distances, the
 potential simply goes over smoothly to the effective potential for the elementary field $\phi$, with the
 replacement $\chi\rightarrow\phi^{2}$. From (\ref{eq:exploop1}) the leading correction for small $\rho$ 
 is evidently 
\begin{equation}
\label{eq:correction}
  \delta_{K}V_{\rm eff}=-\frac{1}{4}(m_{R}^{2}+\lambda_{R}\chi)\int\frac{d^{3}p}{(2\pi)^{3}}\frac{1}{\sqrt{\vec{p}^2+m_{R}^{2}+3\lambda_{R}\chi}}\tilde{K}(\vec{p})
\end{equation}
 which is clearly of order $\hbar^{2}$ at the extremum (where $m_{R}^{2}+\lambda_{R}\chi\sim O(\hbar)$), showing that the extremal energy is independent of
 the choice of smearing function to this order.  The actual value of the composite field in the ground
 state is shifted by the smearing. The amount of the shift  (extremizing $V$ to leading order in $\hbar$)
 is easily seen to be
\begin{equation}
\label{eq:chishift}
 \delta_{K}\chi=\frac{1}{2}\int\frac{d^{3}p}{(2\pi)^{3}}\frac{1}{\sqrt{\vec{p}^2+m_{\rm ph}^{2}}}\tilde{K}(\vec{p})
\end{equation}
where $m_{\rm ph}=m_{R}^{2}+3\lambda_{R}\chi_{\rm ex}$ is the physical scalar mass ($\chi_{\rm ex}$ is
 the extremum value of the composite field).  
  To understand the origin of this shift  we can write
\bdm
 <0| \int d\vec{r}\phi(x+\vec{r}/2)K(\vec{r})\phi(x-\vec{r}/2)|0> =<\phi>^{2}+\int d\vec{r}K(\vec{r})\Delta_{F}(\vec{r})
\edm
 The second term is just the connected contribution to the expectation value of the composite bilocal 
 operator, i.e the propagator for the physical mode smeared with $K$. Going to momentum space and
 integrating out the energy component, we find
\bdm
\label{eq:chishift2}
\int d\vec{r}K(\vec{r})\Delta_{F}(\vec{r})=\frac{1}{2}\int\frac{d^{3}p}{(2\pi)^{3}}\tilde{K}(\vec{p})\frac{1}{\sqrt{\vec{p}^{2}+m_{\rm ph}^{2}}}
\edm

 Thus  we see that the shift in $<\chi>$ at the minimum $\chi=-\frac{m_{R}^{2}}{\lambda_{R}}$ 
 induced by the smearing  is exactly equal to  (\ref{eq:chishift}). In general, the composite field can be
 interpreted as equivalent to the square of the usual elementary order parameter as long as the smearing
 is carried out  over scales much larger than the inverse mass gap of the physical modes. The above argument
 also makes it clear that the domain of definition of the effective potential, for a fixed finite smearing scale
 $\rho$, actually begins at $\chi=\chi_{\rm min}>0$ where $\chi_{\rm min}$ is, from the Lehmann 
 representation, at least as large as $z\int\frac{d^{3}p}{(2\pi)^{3}}\tilde{K}(\vec{p})\frac{1}{2\sqrt{p^{2}+m_{\rm ph}^{2}}}$, $z$ is the (finite) residue of the renormalized $\phi$ propagator at the physical pole $p^2=
 m_{\rm ph}^{2}$. For $\rho$ small, this end-point can be made as close as we wish to zero.

 The upshot of the preceding discussion is simply this: the smearing dependence of the effective potential
 defined with a smeared composite field  will be exponentially small as long as the scale of smearing is
 kept large compared to the Compton wavelength of the physical scalar.  Once the field points are
  sufficiently far apart, the expectation of the composite field reverts by clustering to the square of the 
 $vev$ of the elementary field, with exponentially small corrections arising from the connected part of the
 bilocal  expectation value.  This means that ambiguities arising from various choices of the bilocal
 smearing are (a) physically well understood, and (b)  reducible to  an arbitrarily small value.

\section{Composite Effective Potential for the Higgs-Abelian model- Coulomb Gauge}
 In this section we shall repeat the calculation of the 1-loop composite effective potential in a
 physically more interesting case, that of a spontaneously broken gauge symmetry. As we 
 wish to work with gauge-invariant quantities while retaining a strict energy interpretation for
 the potential, we shall use a gauge-invariant bilocal operator as our probe of symmetry
 breaking, but work in a physical gauge where a positive Hamiltonian and positive metric Hilbert space
 obtain. In such a situation the standard argument \cite{Zinn} shows that the value of
 the effective potential at any field value is the minimum energy compatible with that
 value {\em in a physical state}  (with the usual proviso that the effective potential is the 
 convex hull of the perturbatively computed Legendre transform). 
Either Coulomb or axial gauge would be suitable for this purpose. The calculation is 
 somewhat simpler in Coulomb gauge, where we can maintain the smooth smearing procedure
 used in Section 2 for the discrete symmetry case, so we shall use this gauge henceforth.

   The Lagrangian is a Higgs-abelian model, with an additional fermion (the top quark, in the electroweak 
 context) coupled chirally as in  the sigma-model:
\begin{equation}
\label{eq:higgslagr}
 {\cal L}= |(\partial_{\mu}+ieA_{\mu})\phi|^{2}-\frac{1}{4}F_{\mu\nu}^{2}-m^{2}\phi^{*}\phi-P(\phi^{*}\phi)
+\bar{\psi}i (\sl{\partial} - i {e \over 2} \gamma_5 \sl{A})\psi-g_{y}\bar{\psi}(\phi_{1}+i\gamma_{5}\phi_{2})\psi
\end{equation}
 where the scalar field $\phi=\frac{1}{\sqrt{2}}(\phi_{1}+i\phi_{2})$ is now complex.  The bilocal operator 
 used to define a potential will now be taken as
\begin{equation}
\label{eq:bilocgi}
  \chi(t,\vec{x})\equiv \int d\vec{r}\phi^{*}(t,\vec{x}+\vec{r}/2)K(\vec{r})e^{i\int d^{3}y \vec{{\cal J}}(\vec{x},\vec{y},\vec{r})\cdot\vec{A}(t,\vec{y})}\phi(t,\vec{x}-\vec{r}/2)
\end{equation}
 where the c-number current $\vec{{\cal J}}(\vec{x},\vec{y},\vec{r})$ satisfies
\begin{eqnarray}
 \vec{\nabla}_{\vec{y}}\cdot\vec{{\cal J}}(\vec{x},\vec{y},\vec{r})&=&\delta(\vec{x} - \vec{y}-\vec{r}/2)-\delta(\vec{x}-\vec{y}+\vec{r}/2) \\
   \vec{{\cal J}}(\vec{x},\vec{y},\vec{r}) &=& \vec{\nabla}_{\vec{y}}\sigma
\end{eqnarray}
 Here $\sigma(x)$ is a c-number scalar field which is formally identical to the electrostatic potential in a dipole
 field (where $\vec{{\cal J}}$ corresponds to an electric dipole field).  The choice of $\vec{{\cal J}}$ ensures
 that $\chi$ is a gauge-invariant field. Moreover, the gauge-invariantizing phase factor vanishes in Coulomb
 gauge, by a spatial integration by parts in the exponent. Also, the composite field $\chi$ contains by explicit
 construction only fields on a single time-slice, so the Schr\"odinger picture essential for the energy interpretation
 of the potential is maintained. Rigorous arguments show  \cite{KK}
 that the bilocal operator with a smeared gauge-invariantizing string of this type is a valid order parameter
  for SSB in this theory, as the expectation value of this bilocal operator remains nonzero
 even when the field points are taken infinitely far apart.
Finally, insertion of a complete set of states (in Coulomb gauge we have a
 positive metric Hilbert space) leads to positivity of $<0|\chi|0>$ as for the simple theory of Section 2 provided
 the Fourier transform of the smearing function $K(\vec{r})$ is positive.  In a slight change of notation from the preceding
 section, henceforth bare masses and couplings will be unsubscripted, and we no longer display explicitly 
 powers of $\hbar$. 

     As we are dealing with an abelian theory here, the functional integral over the gauge degrees of  freedom
 is gaussian, and it will be convenient to perform this integration (in Coulomb gauge) immediately.  The integration
 over the fermion field may likewise be performed at the outset. The fermion
 determinant gives a contribution to the effective action which is explicitly
 of order $\hbar$, and which is a functional of the scalar and gauge fields. The
 gauge field dependence of this functional only contributes at order
 $\hbar^{2}$, and may therefore be neglected when the gauge field integrations
 are performed. (The scalar field dependence of the fermionic determinant will be
 computed and included below.) The 
 remaining functional integral over the scalar field will then be evaluated using
 a saddle-point expansion as in 
 Section 2. In doing the gauge integrations, it is best to separate the transverse and Coulomb modes, which
 give rise to physically distinct  contributions. Thus, defining
\begin{eqnarray}
\label{eq:coulombdefs}
  J^{\mu}&\equiv& ie(\phi^{*}\partial^{\mu}\phi-\phi\partial^{\mu}\phi^{*}) \\
  D &\equiv& \Box+2e^{2}\phi^{*}\phi
\end{eqnarray}
 we find for the integral over the transverse modes
\begin{eqnarray}
\label{eq:transversez}
  -i\ln{(Z_{\rm tr})} &\equiv&-i\ln{\int{\cal D}\vec{A}\delta(\vec{\nabla}\cdot\vec{A})e^{i\int(-\frac{1}{2}\vec{A}
D\vec{A}+\vec{J}\cdot\vec{A})d^{4}x}} \\
  &=& \frac{1}{2}\int d^{4}x J_{i}D^{-1/2}(\delta_{ij}-\frac{\partial_{i}\partial_{j}} {\Delta})D^{-1/2}J_{j}+i{\rm tr}\ln{D}
\end{eqnarray}
 while the integral over the Coulomb mode yields
\begin{eqnarray}
\label{eq:coulombz}
 -i\ln{(Z_{c})}&\equiv&-i\ln{\int{\cal D}A_{0}e^{i\int(\frac{1}{2}A_{0}(-\Delta+2e^{2}\phi^{*}\phi)A_{0}-J_{0}A_{0})d^{4}x}} \\
 &=&-\frac{1}{2}\int d^{4}x J_{0}\frac{1}{-\Delta+2e^{2}\phi^{*}\phi}J_{0}+\frac{i}{2}{\rm tr}\ln{(-\Delta+2e^{2}\phi^{*}\phi)}
\end{eqnarray}
 Note that terms quadratic in $J$ are effectively tree-level in powers of Planck's constant, and must therefore
 be included in the effective scalar action when we perform the saddle-point evaluation of the effective potential.

   After integrating out the gauge degrees of freedom, the effective tree-level scalar Lagrangian, augmented by a source
 term for the bilocal $\chi$ field, becomes 
\begin{eqnarray}
\label{eq:efflagr}
 {\cal L}^{\phi}_{\rm eff}&=&\partial_{\mu}\phi^{*}\partial^{\mu}\phi-\phi^{*}\tilde{m}^{2}\phi-P(\phi^{*}\phi)
 +\frac{1}{2}J_{i}(\phi)D^{-1/2}(\delta_{ij}-\frac{\partial_{i}\partial_{j}} {\Delta})D^{-1/2}J_{j}(\phi) \nonumber \\
& -&\frac{1}{2}J_{0}(\phi)\frac{1}{-\Delta+2e^{2}\phi^{*}\phi}J_{0}(\phi)
\end{eqnarray}
 where $\tilde{m}^{2}\equiv m^{2}+jK$ incorporates the source bilinear in the field together with the bare mass
 term into a single nonlocal kernel. Now we expand $\phi = \phi_{0}+\hat{\phi}$, where $\frac{\partial {\cal L}^{\phi}_{\rm eff}}{\partial\phi}(\phi=\phi_{0})=0$ and take the source $j(x)$ and consequently the response $\phi_{0}(x)=\phi_{0}$
 constant. One finds
\begin{equation}
\label{eq:quadlag}
 {\cal L}^{\phi}_{\rm eff} (\phi) \simeq   {\cal L}^{\phi}_{\rm eff} (\phi_{0})+\frac{1}{2}(\hat{\phi}^{*}\;\;\hat{\phi}){\cal M}_{\phi}
\left(\begin{array}{c}\hat{\phi}^{*} \\ \hat{\phi}\end{array} \right)
\end{equation}
 Note that only the Coulomb part of (\ref{eq:efflagr}) contributes in the translationally invariant limit to the
 quadratic form ${\cal M}_{\phi}$, as in this limit $J_{i}\rightarrow ie(\phi_{0}^{*}\partial_{i}\hat{\phi}-\phi_{0}\partial_{i}
\hat{\phi}^{*})$ and $(\delta_{ij}-\frac{\partial_{i}\partial_{j}}{\Delta})\partial_{j}\hat{\phi}=0$. Explicitly,\\
$ {\cal M}_{\phi}= $ \\
\[ \left( \begin{array}{cc}
  -\frac{\partial^{2}P}{\partial\phi^{*2}}|_{\phi_{0}}-e^{2}\phi_{0}^{2}\frac{\partial_{0}^{2}}{-\Delta+2e^{2}|\phi_{0}|^{2}} & -(\Box+\tilde{m}^{2})-\frac{\partial^{2}P}{\partial\phi\partial\phi^{*}}|_{\phi_{0}}+e^{2}|\phi_{0}|^{2}
\frac{\partial_{0}^{2}}{-\Delta+2e^{2}|\phi_{0}|^{2}}\\
-(\Box+\tilde{m}^{2})-\frac{\partial^{2}P}{\partial\phi\partial\phi^{*}}|_{\phi_{0}}+e^{2}|\phi_{0}|^{2}
\frac{\partial_{0}^{2}}{-\Delta+2e^{2}|\phi_{0}|^{2}}& -\frac{\partial^{2}P}{\partial\phi^{2}}|_{\phi_{0}}-e^{2}\phi_{0}^
{*2}\frac{\partial_{0}^{2}}{-\Delta+2e^{2}|\phi_{0}|^{2}} \end{array} \right) \]

   Specializing to the case $P(\phi)=\lambda(\phi^{*}\phi)^{2}$,  and noting that in the translationally
 invariant limit $\chi=\int d^{4}x \, d\vec{r}\phi^{*}(x-\vec{r}/2)K(\vec{r})\phi(x+\vec{r}/2)=|\phi_{0}|^{2}$,
this simplifies to \\
$ {\cal M}_{\phi}=$ \\
\[ \left( \begin{array}{cc}
 -(2\lambda+e^{2}\frac{\partial_{0}^{2}}{-\Delta+2e^{2}\chi} )\phi_{0}^{2}& -(\Box+\tilde{m}^{2}+6\lambda\chi)+(2\lambda+e^{2}
\frac{\partial_{0}^{2}}{-\Delta+2e^{2}\chi})\chi \\
-(\Box+\tilde{m}^{2}+6\lambda\chi)+(2\lambda+e^{2}
\frac{\partial_{0}^{2}}{-\Delta+2e^{2}\chi})\chi & -(2\lambda+e^{2}
\frac{\partial_{0}^{2}}{-\Delta+2e^{2}\chi})\phi_{0}^{*2} \end{array} \right) \]
whence the integral over the fluctuation fields $\hat{\phi}$ yields directly
\begin{eqnarray}
\label{eq:detmphi}
\ln{{\rm det}({\cal M}_{\phi})}&=&\ln{{\rm det}\{(\Box+\tilde{m}^{2}+6\lambda\chi)^{2}-
(\Box+\tilde{m}^{2}+6\lambda\chi)
(4\lambda\chi+2e^{2}\chi\frac{\partial_{0}^{2}}{-\Delta+2e^{2}\chi})\}}  \nonumber  \\
&=&\ln{{\rm det}(\Box+\tilde{m}^{2}+6\lambda\chi)}+\ln{{\rm det}\{\Box+\tilde{m}^{2}+2\lambda\chi
 -2e^{2}\chi\frac{\partial_{0}^{2}}{-\Delta+2e^{2}\chi}\}} \nonumber \\
&=& \ln{{\rm det}(\Box+\tilde{m}^{2}+6\lambda\chi)}+\ln{{\rm det}(\Box+2e^{2}\chi)}
-\ln{{\rm det}(-\Delta+2e^{2}\chi)} \nonumber \\
&+&\ln{{\rm det}\{1+(\frac{-\Delta+2e^{2}\chi}{-\Delta})\frac{1}{\Box+2e^{2}\chi}(\tilde{m}^{2}+2\lambda\chi)\}}
+{\rm constant}
\end{eqnarray}
Adding in the one-loop contributions from the gauge integrations we find
\begin{eqnarray}
\label{eq:gaugescalar}
-V_{\rm gauge+scalar}^{\rm 1-loop}&=&i\frac{3}{2}{\rm tr}\ln{(\Box+2e^{2}\chi)}+i\frac{1}{2}{\rm tr}\ln{(\Box+\tilde{m}^{2}
+6\lambda\chi)}  \nonumber \\
 &+&i\frac{1}{2}{\rm tr}\ln{\{1+(\frac{-\Delta+2e^{2}\chi}{-\Delta})\frac{1}{\Box+2e^{2}\chi}(\tilde{m}^{2}+2\lambda\chi)\}}
\end{eqnarray}
 The unpleasant term $\ln{{\rm det}(-\Delta+2e^{2}\chi)}$, which contains a UV divergence of the
 form $\int dp_{0}$ not present in the counterterms, has cancelled, and the remaining terms are
 easily seen to correspond in the broken symmetry phase to three massive gauge vector modes
 and a single massive scalar mode. The complicated and  peculiar last term is a remnant of the 
 long-range Coulomb interaction, and contains divergences which will be removed by the
 counterterms of the theory, as we shall see shortly. In these one-loop contributions, the
 source augmented mass $\tilde{m}$ contains the source $j_{0}$ of the composite field $\chi$,
 determined at tree level from $\frac{\partial{\cal L}^{\phi}_{\rm eff}}{\partial\phi}|_{\phi_{0}}=0$
 for $\phi_{0}$ constant, so
\begin{eqnarray}
\label{eq:source}
 j_{0}&=&(-m^{2}+V^{\prime}(\chi))=-(m^{2}+2\lambda\chi) \\
 \tilde{m}^{2} &=& m^{2}+j_{0}K=m^{2}(1-K)-2\lambda\chi K
\end{eqnarray}
  Integrating out the fermion field gives a similar contribution (except for the characteristic change
 of sign)
\begin{equation}
\label{eq:fermion}
 -V_{\rm fermion}^{\rm 1-loop}=-i\;{\rm tr}\ln{(i \sl{\partial}-g_{y}(\phi_{1}+i\gamma_{5}\phi_{2}))}
\end{equation}
 The full effective potential through 1-loop is thus given by
\begin{equation}
\label{eq:vfull}
  V(\chi) = m^{2}\chi+\lambda\chi^{2}+V_{\rm gauge+scalar}^{\rm 1-loop}+V_{\rm fermion}^{\rm 1-loop}
\end{equation}
 This  expression is as yet unrenormalized- we must rewrite the bare parameters $m,\lambda,g_y$ in terms
 of renormalized parameters and rescale the fields appropriately, at which point the divergences in
 the one-loop contributions will be seen to cancel completely. This calculation will be performed in the
 next section.

\section{Renormalized Composite Effective Potential for the Higgs-Abelian Model}

   To renormalize the result (\ref{eq:vfull}) obtained above,  one may integrate out
 the energy component $p_{0}$ in momentum space and then dimensionally 
 regularize the resulting purely spatial integrals (which are then carried out in
 3-$\epsilon$ dimensions). For example the contribution from the massive gauge vector
 loop was found to be
\begin{eqnarray}
\label{eq:apiece}
 V_{A} &=& -i\frac{3}{2}{\rm tr}\ln{(\Box+2e^{2}\chi)}    \nonumber \\
            &=& -i\frac{3}{2}\int\frac{d^{4}p}{(2\pi)^{4}}\ln{(-p^{2}+2e^{2}\chi)}\;\;\; {\rm Minkowski} \nonumber  \\
             &=&{\rm const}+\frac{3}{2}\int\frac{d^{4}p}{(2\pi)^{4}}\ln{(1+\frac{2e^{2}\chi}{p_{4}^{2}+\vec{p}^{2}})}
\;\;\;{\rm Euclidean} \nonumber \\
   &\rightarrow&\frac{3-\epsilon}{2}\int\frac{d^{3-\epsilon}p}{(2\pi)^{3-\epsilon}}(\sqrt{\vec{p}^{2}+2e^{2}\chi}
-|\vec{p}|)
\end{eqnarray}
where the final spatial integral, together with the prefactor ($g^{\mu}_{\mu}-1$) counting spatial gauge modes, has been
 dimensionally continued.  The  divergent part of $V_{A}$ can be separated off as follows:
\begin{eqnarray}
\label{eq:vasep}
 V_{A}&=&\frac{3-\epsilon}{2}\int\frac{d^{3-\epsilon}p}{(2\pi)^{3-\epsilon}}(\sqrt{\vec{p}^{2}+2e^{2}\chi}
-|\vec{p}|-\frac{e^{2}\chi}{|\vec{p}|}+\frac{1}{2}\frac{e^{4}\chi^{2}}{(|\vec{p}|^{2}+\eta^{2})
^{3/2}}) +V_{A,\rm div} \nonumber \\
 V_{A,\rm div}&\equiv&-\frac{3-\epsilon}{4}\int\frac{d^{3-\epsilon}p}{(2\pi)^{3-\epsilon}}\frac{e^{4}\chi^{2}}{(|\vec{p}|^{2}+\eta^{2})
^{3/2}}  \nonumber  \\
  &=&-\frac{3}{16\pi^{2}}e^{4}\chi^{2}(\Delta_{\epsilon}-\ln{(\frac{\eta^{2}}{\mu^{2}})}-\frac{2}{3})
\end{eqnarray}
 Here the divergent piece $\Delta_{\epsilon}$ is defined in (\ref{eq:counterterm}), and $\eta$
 is an infrared cutoff which will be set to zero at the end. The first term in (\ref{eq:vasep})
 has all the subtractions needed to ensure that it remains finite as $\epsilon\rightarrow 0$
 so we may define a $\overline{\rm MS}$-renormalized gauge boson contribution (obtained as usual
 by subtracting the part proportional to $\Delta_{\epsilon}$ in $V_{A,\rm div}$) as
\begin{equation}
\label{eq:varen}
 V_{A,\rm ren}\equiv\frac{3}{2}\int\frac{d^{3}p}{(2\pi)^{3}}(\sqrt{\vec{p}^{2}+2e^{2}\chi}-|\vec{p}|
 -\frac{e^{2}\chi}{|\vec{p}|}+\frac{1}{2}\frac{e^{4}\chi^{2}}{(\vec{p}^{2}+\eta^{2})^{3/2}})
 +\frac{3}{16\pi^{2}}e^{4}\chi^{2}(\ln{(\frac{\eta^{2}}{\mu^{2}})}+\frac{2}{3})
\end{equation} 
 
  Similar manipulations can be used to extract the singular parts of the massive Higgs, residual
 Coulomb, and fermion contributions to the 1-loop composite effective potential. One obtains
 for the massive scalar loop contribution
\begin{eqnarray}
\label{eq:vbsep}
 V_{B}&=&\frac{1}{2}\int\frac{d^{3-\epsilon}p}{(2\pi)^{3-\epsilon}}(\sqrt{\vec{p}^{2}
+4\lambda\chi+(m^{2}+2\lambda\chi)(1-\tilde{K}(\vec{p}))}-|\vec{p}|) \\
 &=& V_{B,\rm ren}-\frac{1}{64\pi^{2}}(m^{2}+6\lambda\chi)^{2}\Delta_{\epsilon}
\end{eqnarray}
with 
\begin{eqnarray}
\label{eq:vbren}
V_{B,\rm ren}&=&\frac{1}{2}\int\frac{d^{3}p}{(2\pi)^{3}}(\sqrt{\vec{p}^{2}+4\lambda\chi
 +(m^{2}+2\lambda\chi)(1-\tilde{K}(\vec{p}))}-|\vec{p}|-\frac{1}{2|\vec{p}|}(m^{2}+6\lambda\chi) \nonumber \\
&+&\frac{1}{8}\frac{(m^{2}+6\lambda\chi)^{2}}{(\vec{p}^{2}+\eta^{2})^{3/2}})+\frac{1}{64\pi^{2}}
 (m^{2}+6\lambda\chi)^{2}\ln{(\frac{\eta^{2}}{\mu^{2}})}
\end{eqnarray}
 The residual Coulomb contribution also yields a divergent contribution:
\begin{eqnarray}
\label{eq:vcsep}
 V_{C}&=&\frac{1}{2}\int\frac{d^{3-\epsilon}p}{(2\pi)^{3-\epsilon}}\sqrt{\vec{p}^{2}+2e^{2}\chi}
(\sqrt{1+(m^{2}+2\lambda\chi)\frac{1-\tilde{K}(\vec{p})}{\vec{p}^{2}}}-1) \\
 &=&V_{C,\rm ren}+\frac{1}{16\pi^{2}}e^{2}\chi(m^{2}+2\lambda\chi)\Delta_{\epsilon}
-\frac{1}{64\pi^{2}}(m^{2}+2\lambda\chi)^{2}\Delta_{\epsilon}
\end{eqnarray}
 This term couples the Higgs and gauge sectors and the renormalized contribution is
 rather complicated. One finds
\begin{eqnarray}
\label{eq:vcren}
 V_{C,\rm ren}=\frac{1}{2}\int\frac{d^{3}p}{(2\pi)^{3}}&\{&(\sqrt{\vec{p}^{2}+2e^{2}\chi}-
|\vec{p}|-\frac{e^{2}\chi}{|\vec{p}|})(\sqrt{1+(m^{2}+2\lambda\chi)\frac{1-\tilde{K}(\vec{p})}{
\vec{p}^{2}}}-1) \nonumber \\
 &+&|\vec{p}|(\sqrt{1+(m^{2}+2\lambda\chi)\frac{1-\tilde{K}(\vec{p})}{
\vec{p}^{2}}}-1-\frac{m^{2}+2\lambda\chi}{2p^{2}}+\frac{(m^{2}+2\lambda\chi)^{2}}{8
(\vec{p}^{2}+\eta^{2})^{2}}) \nonumber \\
 &+&\frac{e^{2}\chi}{|\vec{p}|}(\sqrt{1+(m^{2}+2\lambda\chi)\frac{1-\tilde{K}(\vec{p})}{
\vec{p}^{2}}}-1-\frac{m^{2}+2\lambda\chi}{2(\vec{p}^{2}+\eta^{2})})\} \nonumber  \\
 &-&\frac{1}{16\pi^{2}}e^{2}\chi(m^{2}+2\lambda\chi)(\ln{(\frac{\eta^{2}}{\mu^{2}})}-2+2\ln{2}) 
\nonumber \\
 &+&\frac{1}{64\pi^{2}}(m^{2}+2\lambda\chi)^{2}(\ln{(\frac{\eta^{2}}{\mu^{2}})}-1+2\ln{2})
\end{eqnarray}
  Finally, the 1 loop fermion contribution gives
\begin{eqnarray}
\label{eq:vdsep}
V_{D}&=&-2\int\frac{d^{3-\epsilon}p}{(2\pi)^{3-\epsilon}}(\sqrt{\vec{p}^{2}+2g_{y}^{2}\chi}
-|\vec{p}|) \\
 &=&V_{D,\rm ren}+\frac{1}{4\pi^{2}}g_{y}^{4}\chi^{2}\Delta_{\epsilon}
\end{eqnarray}
where
\begin{eqnarray}
\label{eq:vdren}
V_{D,\rm ren}&=&-2\int\frac{d^{3}p}{(2\pi)^{3}}(\sqrt{\vec{p}^{2}+2g_{y}^{2}\chi}-|\vec{p}|
-\frac{g_{y}^{2}\chi}{2|\vec{p}|}+\frac{1}{8}\frac{g_{y}^{4}\chi^{2}}{(\vec{p}^{2}+\eta^{2})^{3/2}})
\nonumber \\
&-&\frac{1}{4\pi^{2}}g_{y}^{4}\chi^{2}\ln{(\frac{\eta^{2}}{\mu^{2}})}
\end{eqnarray}

   The 1-loop renormalization of the theory (mass, coupling and field rescaling) implies the
 following cutoff dependence for the coefficients of $\chi$ and $\chi^{2}$ in the
 tree contribution to the potential (individual counterterm contributions are listed
 in Appendix A):
\begin{eqnarray}
\label{eq:gaugecounter}
 m^{2}&\rightarrow& m_{R}^{2}(1+\frac{\Delta_{\epsilon}}{16\pi^{2}}(-e_{R}^{2}+4\lambda_{R})) \\
 \lambda&\rightarrow&\lambda_{R}+\frac{\Delta_{\epsilon}}{16\pi^{2}}(-2e_{R}^{2}\lambda_{R}
 +10\lambda_{R}^{2}+3e_{R}^{4}-4g_{y}^{4})
\end{eqnarray}
 When these replacements are made in the unrenormalized potential (\ref{eq:vfull}), one finds
 that all divergences cancel and we are left with an explicitly finite result for the
 composite potential (now in terms of  $\overline{\rm MS}$-renormalized fields, masses and couplings):
\begin{equation}
\label{eq:vfullren}
 V_{\rm ren}(\chi)=m_{R}^{2}\chi+\lambda_{R}\chi^{2}+V_{A,\rm ren}(\chi)+V_{B,\rm ren}(\chi)
 +V_{C,\rm ren}(\chi)+V_{D,\rm ren}(\chi)
\end{equation}
 with the renormalized potentials given explicitly in (\ref{eq:varen},\ref{eq:vbren},\ref{eq:vcren},
\ref{eq:vdren}) with the mass and couplings therein interpreted as renormalized ones.

\section{Analytic Formulae for Sharp Momentum-space Smearing}

As mentioned previously, 
any choice of smearing function $K\left( \vec{r} \right)$ which adequately suppresses short-distance contributions to $ \left\langle \chi\left( x \right) \right\rangle$ could be used in defining the composite operator.
To the extent that different $K\left( \vec{r} \right)$ isolate only the 
longest-distance modes relevant for SSB they should give similar
values for  $ \left\langle \chi\left( x \right) \right\rangle$ and thus 
express the same physics.  In particular, they should unambiguously 
signal the location and depth
of symmetry-breaking extrema of the effective potential.
One simple choice which permits closed-form expressions for the dimensionally-regulated momentum integrals is  
the step function, $ \tilde{K}\left( \vec{p} \right)= \theta\left(\rho- |\vec{p}| \right)$.
With this choice, the $K$-dependent integrals in $V_B$ and $V_C$ can
be written generically (to ${\cal O}\left( \epsilon^0 \right)$) as
\begin{eqnarray}
\int{d^{3-\epsilon}p\ f\left( \tilde{K}(|\vec{p}|) \right) }&=&
\int_{|\vec{p}|>\rho}{ d^{3-\epsilon}p\ f\left( \tilde{K}=0 \right)}+
\int_{|\vec{p}|<\rho}{ d^3 p\ f\left(\tilde{K}=1 \right)}
\nonumber \\
 &=&\int{ d^{3-\epsilon}p\ f\left(0 \right)}+
\int_{|\vec{p}|<\rho}{ d^3 p\ \left(f\left( 1 \right)-f\left( 0 \right) \right)}
\end{eqnarray}
The first integral contains the UV divergence and 
can be evaluated in the
usual manner.  It is independent of the smearing radius $\rho$.
The second integral is manifestly UV finite, contains all of the dependence on
$\rho$, and is also straightforward to evaluate.  Thus we see that at one loop the 
effective potential for the bilocal
composite operator naturally decomposes into
a 
piece which is the elementary field effective potential (written in terms of $\chi=\phi^* \phi$)
and a piece dependent on the smearing scale which introduces no new UV divergences,
unlike the local composite operator $\phi^2(x)$.  

The explicit expressions for terms of the effective potential are: 
\begin{eqnarray}
V_{\rm A}\left( \chi \right)&=&{{3 m_V^4} \over {64 \pi^2}}
\left[ -\Delta_\epsilon +\ln{{{m_V^2} \over {\mu^2}}} - {5 \over 6} \right] \\
& & \nonumber \\
V_{\rm B}\left( \chi \right)&=&V_{\rm B,1}\left( \chi \right)+V_{\rm B,2}\left( \chi \right) \\
V_{\rm B,1}\left( \chi \right)&=&{{m_h^4} \over {64 \pi^2}} 
\left[ -\Delta_\epsilon +\ln{{{m_h^2 } \over {\mu^2}}} - {3 \over 2} \right]  \\
V_{\rm B,2}\left( \chi \right)&=&
{1 \over {64 \pi^2}} 
\left\{2 \rho \sqrt{\rho^2 + 4 \lambda\chi} \left( 2 \rho^2 + 4 \lambda \chi \right)-
\left( 4 \lambda \chi \right)^2
\ln{   {\left(\rho+\sqrt{\rho^2 + 4 \lambda \chi}\right)^2} \over {4 \lambda \chi}   }  \right\}-
\nonumber  \\
& & 
{1 \over {64 \pi^2}} 
\left\{2 \rho \sqrt{\rho^2 + m_h^2} \left( 2 \rho^2 + m_h^2 \right)-m_h^4
\ln{   {\left(\rho+\sqrt{\rho^2 + m_h^2}\right)^2} \over {m_h^2}  }  \right\} \\
& &  \nonumber  \\
V_{\rm C}\left( \chi \right)&=&V_{\rm C,1}\left( \chi \right)+V_{\rm C,2}\left( \chi \right) \\
V_{\rm C,1}\left( \chi \right)&=&{1 \over {64 \pi^2}} 
\Bigg\{ -( m_g^2 -  m_V^2)^2  \left( \Delta_\epsilon + {3 \over 2} \right) 
-2 m_g m_V (m_g-m_V)^2 + \nonumber \\
 & &(m_g^2 - m_V^2)^2 
 \ln{  {\left( m_g+ m_V \right)^2} \over {\mu^2}  } 
\Bigg\} 
 - {m_V^4 \over 64 \pi^2} \left\{     - \left (\Delta_\epsilon + {3 \over 2}  \right)
+ \ln{m_V^2 \over \mu^2}
     \right\} \\
V_{\rm C,2}\left( \chi \right)&=&
{2 \over {64 \pi^2}} 
\left\{  \rho \sqrt{m_g^2 + \rho^2} \left( m_g^2 + 2 \rho^2 \right) - m_g^4 
\ln{{\rho+\sqrt{\rho^2 + m_g^2}} \over {m_g} }\right\}  - \nonumber \\
& & {2 \over {64 \pi^2}} 
\Biggl\{   \sqrt{m_g^2 + \rho^2} \sqrt{m_V^2 +\rho^2}  \left( m_g^2 + m_V^2 +  2 \rho^2 \right) -  m_g m_V (m_g^2 + m_V^2) - \nonumber \\
& &
\left( m_g^2 - m_V^2 \right)^2
\ln{{\sqrt{\rho^2 + m_g^2}+ \sqrt{\rho^2 + m_V^2}} \over {m_g+ m_V} }\Biggr\} \\
& & \nonumber \\
V_{\rm D}\left( \chi \right)&=&-{4 \over {64 \pi^2}} m_f^4 \left[ -\Delta_\epsilon  +\ln{{m_f^2} \over {\mu^2}} - {3 \over 2} \right]
\end{eqnarray}
where $m_g^2=m^2 + 2 \lambda \chi$, $m_V^2=2 e^2 \chi$, $m_h^2=
m^2+ 6 \lambda \chi$ and $m_f^2=2 g_y^2 \chi$.

As noted previously, for $\rho \rightarrow 0$ the $\rho$-dependent
terms go smoothly to zero.  For $\rho \rightarrow \infty$ new quadratic
divergences arise as the point-split composite operator approaches
the local composite operator.  

\section{Renormalization Group Improvement and Vacuum Stability Bounds}

The composite operator effective potential 
formalism developed above provides a 
gauge-invariant framework for obtaining the information about symmetry 
breaking which is partially obscured by gauge dependence and UV problems in other treatments.
While the value of the elementary-field effective potential is gauge invariant at any of its local
extrema, the field value at which the extremum occurs (i.e. the expectation value of the elementary
scalar field in the corresponding phase of the theory) is not.  Thus, associating the
gauge dependent expectation value of the elementary field at which some feature of the
effective potential occurs with gauge invariant physical quantities is a practice of 
questionable validity.  The expectation value of the composite operator is by construction free
of these gauge ambiguities. Because the effective potential of the composite operator has
an energy interpretation, the value of the effective potential at any given value of
 the composite order parameter corresponds to 
the minimum physical energy density compatible with that value.  Consequently, the global minimum of this effective potential
should correspond to the true vacuum of the theory and the value of $\chi$ at which it
occurs should also have physical meaning.

One context in which such a treatment is useful is in the formulation of a lower bound on the 
Standard Model Higgs mass from vacuum stability considerations.  If the 
electroweak vacuum is assumed to be stable (rather than merely metastable) then it must be the 
global minimum of the effective potential (in both the case of the composite 
operator effective potential and of the elementary field effective potential).  In practice, however, we
expect that the Standard Model is the low-energy effective theory of a more fundamental 
high-scale theory and so will only be an accurate description of physics up to some
energy scale (the `new physics' scale, perhaps characterized by the masses of new heavy 
particles).  Thus, it is only consistent to demand that the electroweak vacuum is the global 
minimum up to the scale at which the effective theory breaks down, beyond which the model
is simply no longer accurate.

For the elementary field effective potential, the statement that the value of the 
effective potential is larger than the value at the electroweak minimum up to some
scale $\Lambda$ is a gauge-dependent statement \cite{LW}.
As such it is an unsatisfactory criterion for defining a `new physics' scale.
For the composite operator effective potential defined and evaluated in this
 paper  there is no gauge dependence
and we may interpret the composite operator
expectation value as a physically meaningful energy scale.

Since the Standard Model may be valid up to very high energies it will be necessary to
study the effective potential for field expectation values much larger than the 
electroweak scale.
However, it is well-known that in order to study the effective potential at large field values, the
usual perturbative loop expansion for the effective potential may be inadequate.
For the elementary field effective potential, the loop expansion generates terms generically of form 
${g^4 \phi^4 \over (4 \pi)^2} \ln{g^2 \phi^2 \over \mu^2}$ (where $g$ is any of the couplings).  If the 
couplings $\hat{g}$ and the renormalization scale $\mu$ are given values characteristic of the electroweak 
scale (e.g. $\mu={\rm M_Z}$ and the couplings given their values at the Z-scale),
considering $\phi \gg \mu$ will invalidate the perturbative expansion.  The range
of validity of the approximation may be improved by utilizing the renormalization group 
(RG) improved effective potential \cite{CW, Einhorn}.  Similar considerations apply to the
composite operator effective potential.
  
The full (all-orders) effective potential is independent of $\mu$.    This
independence can be expressed as a first-order differential equation
\bdm
\mu {d \over d \mu}{\vef \left( \chi,m,e,\lambda,g_y,\mu\right)}=0
\edm
This equation can be solved and, combined with dimensional analysis, yields the RG-improved
effective potential  \cite{LW,Einhorn}
\bdm
{\vef\left( s^2 \chi_i,\hat{g}_i,m_i,\mu \right)}=\left[ \zeta\left( s \right) \right]^4\vef\left(\chi_i ,\hat{g}\left( s, \hat{g}_i \right), m \left( s, m_i \right), \mu \right)
\edm
where
\bdm
\gamma_{\phi}=
{1 \over 2} {\mu \over {Z_{\chi}}} {{d Z_{\chi}} \over {d \mu}}
\edm
 Note that $\gamma_{\phi}$ in (69) is really one half of the anomalous dimension of the
 gauge-invariant composite operator $\chi$. In Coulomb gauge (and only in Coulomb gauge) the
 gauge-invariantizing string factor in $\chi$ is equal to unity and the operator therefore renormalizes
 as the square of the elementary field in this gauge. Accordingly $\gamma_{\phi}$ is just the
 usual anomalous dimension of the elementary scalar in Coulomb gauge. In other gauges, there
 will be a divergent contribution from a nontrivial string factor which compensates for the
 gauge-variance of $\gamma_{\phi}$. The scale factor in (68) is just
\bdm
\zeta\left( s \right)=\exp{\left[\int_{0}^{\log{s}}{{1 \over {\gamma_{\phi}\left( x \right)+1}}  dx} \right]}
\edm
 Here $s^2={{\chi} \over {\chi_i} }$ and  $\hat{g}$ represents the set of couplings $g_y, \lambda$, and $e$.
The running couplings $\hat{g}$ are solutions to the equations
\bea
{d \lambda(s)\over ds}={\overline \beta}_\lambda(\hat{g}(s))= {\beta_\lambda \over 1+ \gamma_\phi} &, & \lambda(0)=\lambda_i
\eea
and analogous for the other couplings.  The $\hat{g}_i$ are the values for the couplings
at the initial scale $\chi_i$, which we will take to be around the electroweak scale.  The $\beta$ 
functions can be computed (in a loop expansion) from the $\msb$ counterterms 
of the theory and are listed in Appendix A.
Whereas the unimproved effective potential was trustworthy only for $\chi$ 
such that ${g_i^2 \over (4 \pi)^2} \ln{g_i^2 s^2} \ll 1$, the 
RG improved effective potential will be reliable as long as the 
running couplings remain small.  

It has been demonstrated (in the context of the elementary field effective potential) that the $n$-loop effective
potential improved using $n+1$ loop $\beta$ and $\gamma$ functions resums the $n^{\rm th}$-to-leading
logs \cite{Bando, Kastening}.  Since we are demonstrating the utility of a calculational tool rather than
pursuing a precise numerical result, we will be satisfied to sum the leading logs only.  It is thus sufficient to consider the one-loop effective potential with the tree-level piece run with one-loop $\beta$ functions. 
In the large-field ($\chi \gg m^2$) limit the RG improved effective potential can be written \cite{CEQ1,CEQ2,CEQ3}
\bdm
{\vef\left( s^2 \chi_i,\hat{g}_i,\mu \right)}= 
\lef\left(\chi_i, \hat{g}\left( s,\hat{g}_i \right), \mu \right)
\left[ \chi_i \zeta^2\left( s \right) \right]^2
\edm
\bdm
\lef(s)=\lambda(s)+ \Delta\lambda(\hat{g}_i,\chi_i,\mu)
\edm

We will demonstrate the calculation of the vacuum stability bound using the RG-improved composite operator effective potential in the context of a toy model, the abelian Higgs model coupled to a fermion.
The issue of  vacuum stability arises in qualitatively the same way in the abelian Higgs+fermion model as  in the Standard Model, as 
a result of the large negative contribution of top quark loops to $\beta_\lambda$.
For some electroweak-scale boundary condition on $\lambda(s)$, $\lambda(s_{EW})=\lambda_i$, the top quark term in $\beta_\lambda$ drives $\lef(s)$ negative for sufficiently large $s$.  For large $s$ the tree-level RG-improved effective potential $\vef(\chi_i,s) \approx \lef(s) \chi_i^2 \zeta(s)^4 $ then falls rapidly below $\vef(s_{EW})$ and the electroweak vacuum is
unstable.  We define $s_{VI}$ as the value of $s$ at which
 at which the RG improved effective potential equals zero.  
Since $\zeta(s) > 0$,  $\vef(s_{VI})=0$ is equivalent to $\lef(s_{VI})=0$.
For the electroweak vacuum to remain the global minimum of the theory, the `new physics' must enter before $s_{VI}$ (however, see \cite{HungSher} for qualification).  For a fixed $s_{VI}$ there corresponds a minimum $\lambda_i$
below which the electroweak vacuum is destabilized too early.  This then translates
to a lower bound on the Higgs pole mass.  

\section{Numerical Results}

\begin{figure}
\centering
\epsfysize=4in  
\hspace*{0in}
\epsffile{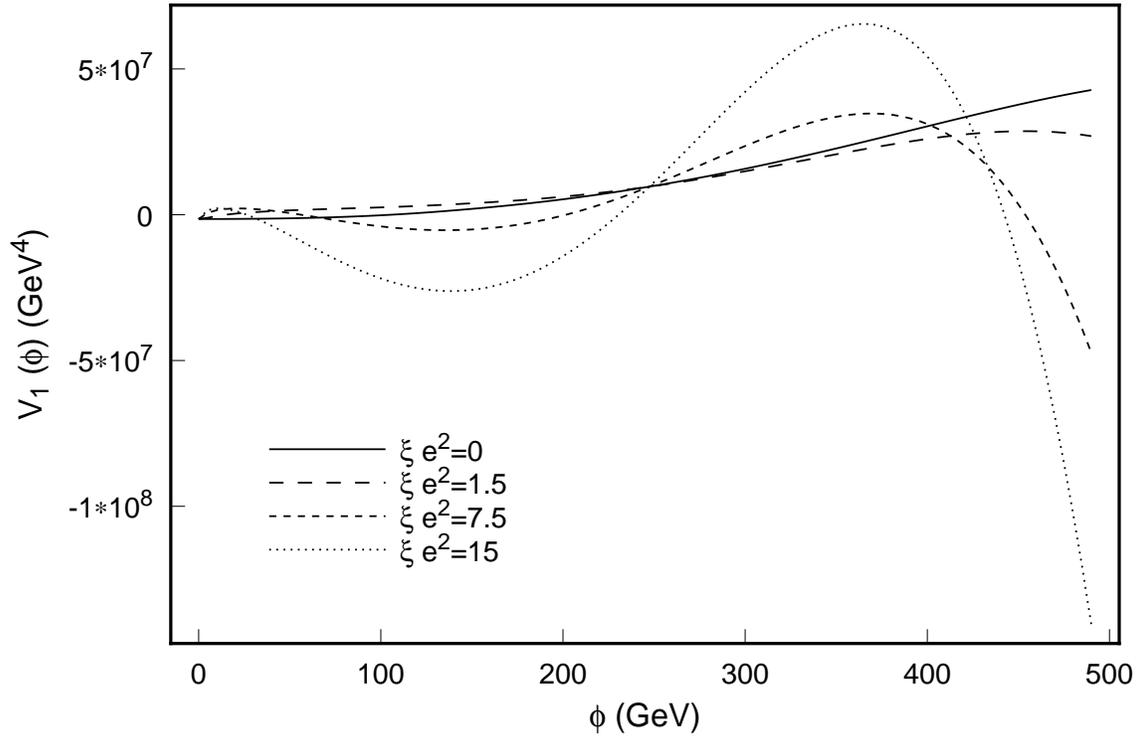} 
\caption{One-loop corrections to the effective potentials in the $\rxi$ gauge for several values of the gauge parameter. }
\label{variousxi_V1} 
\end{figure}

\begin{figure}
\centering
\epsfysize=4in  
\hspace*{0in}
\epsffile{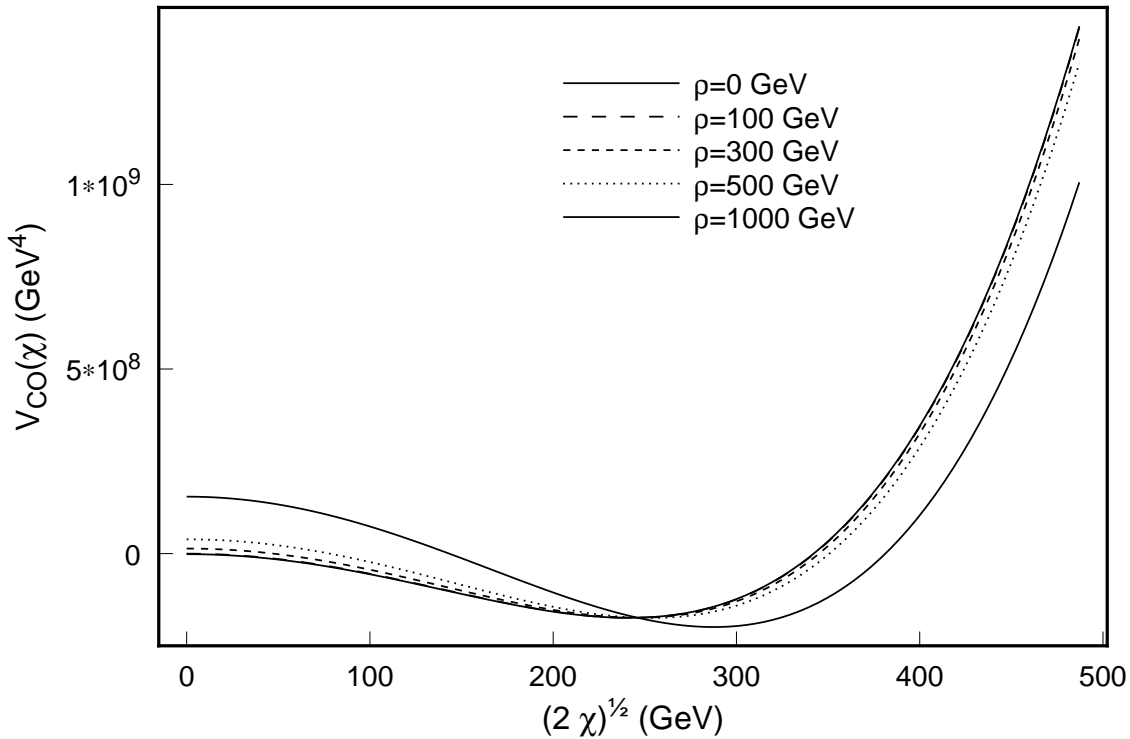} 
\caption{Effective potentials for the composite operator using a sharp $\vec{p}$-space cutoff smearing function.  For cutoff momenta of order the mass parameter ($m=110 \, {\rm GeV}$) or smaller, the curves are essentially indistinguishable.}
\label{unimprovedvariousa} 
\end{figure}

\begin{figure}
\centering
\epsfysize=4in  
\hspace*{0in}
\epsffile{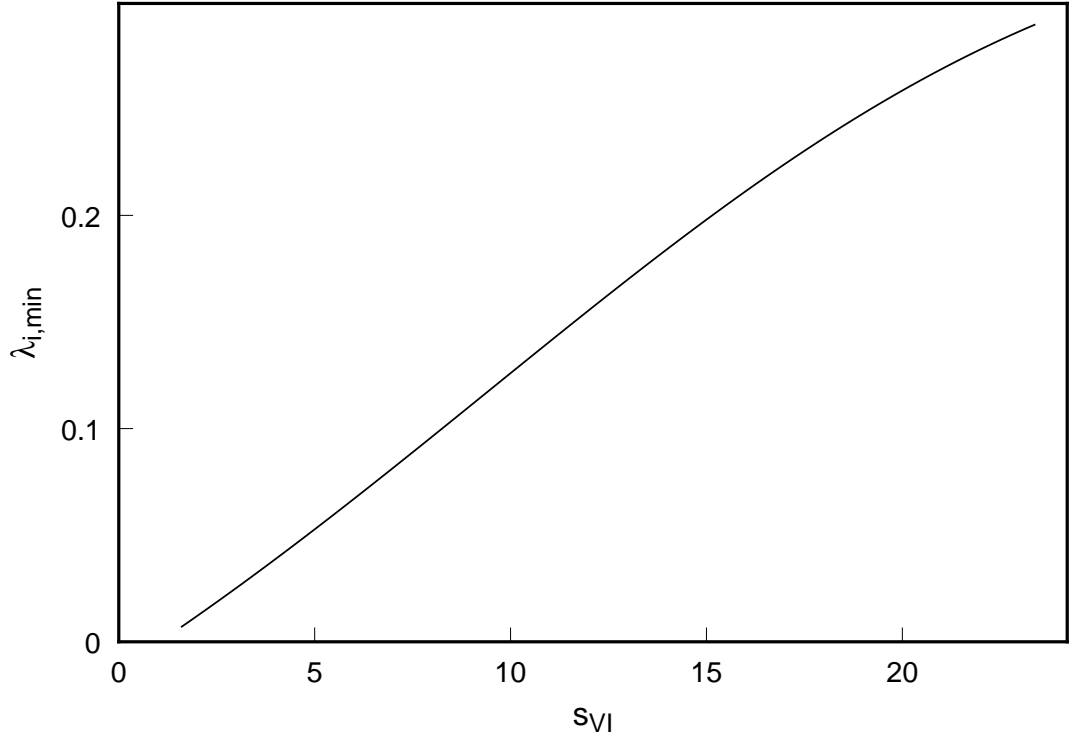} 
\caption{$\lambda_{i,{\rm min}}$ vs. $s_{VI}$ for the composite operator effective potential with $\rho=0$.  The corresponding curve obtained using the elementary-field effective potential in Landau gauge is virtually indistinguishable.  In this plot, $e_i^2=0.15, g_y^2=0.5, v=246 \, {\rm GeV}, \mu^2=\chi_i= {1 \over 2} v^2$.}
\label{lambdai_vs_sVI} 
\end{figure}

\begin{figure}
\centering
\epsfysize=4in  
\hspace*{0in}
\epsffile{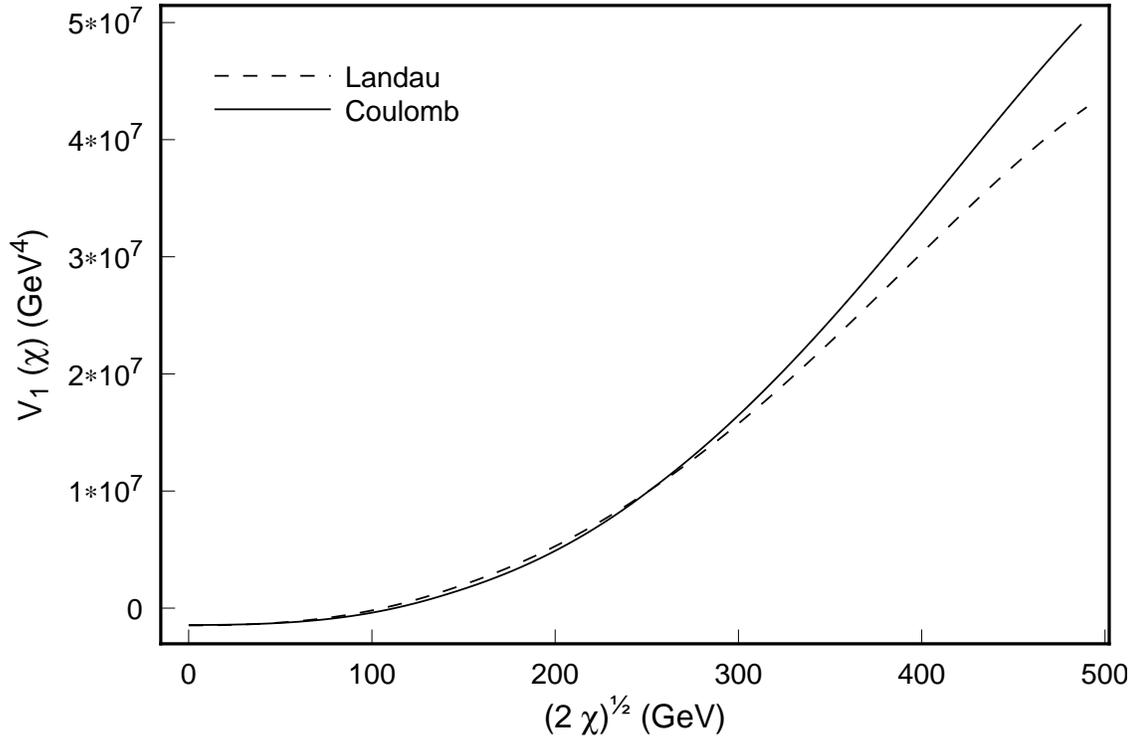} 
\caption{One-loop corrections to the elementary field effective potential in Landau gauge 
and composite operator effective potential (for cutoff momentum zero), for $g^2=0.15$, $g_y^2=0.5$, $\lambda=0.2$, $\mu=v$.  Both curves are without RG improvement.}
\label{Landau_Coulomb_V1} 
\end{figure}

We present here some numerical results to demonstrate the qualitative features of the composite operator
effective potential and its use in the study of vacuum stability bounds.  A set of initial values of 
parameters has been chosen ($e_i^2=0.15, g_y^2=0.5, v=246 \, {\rm GeV}, \mu=v$) to resemble the Standard Model,
but these plots should not be construed as a serious attempt to calculate Standard Model quantities.
\begin{figure}
\centering
\epsfysize=4in  
\hspace*{0in}
\epsffile{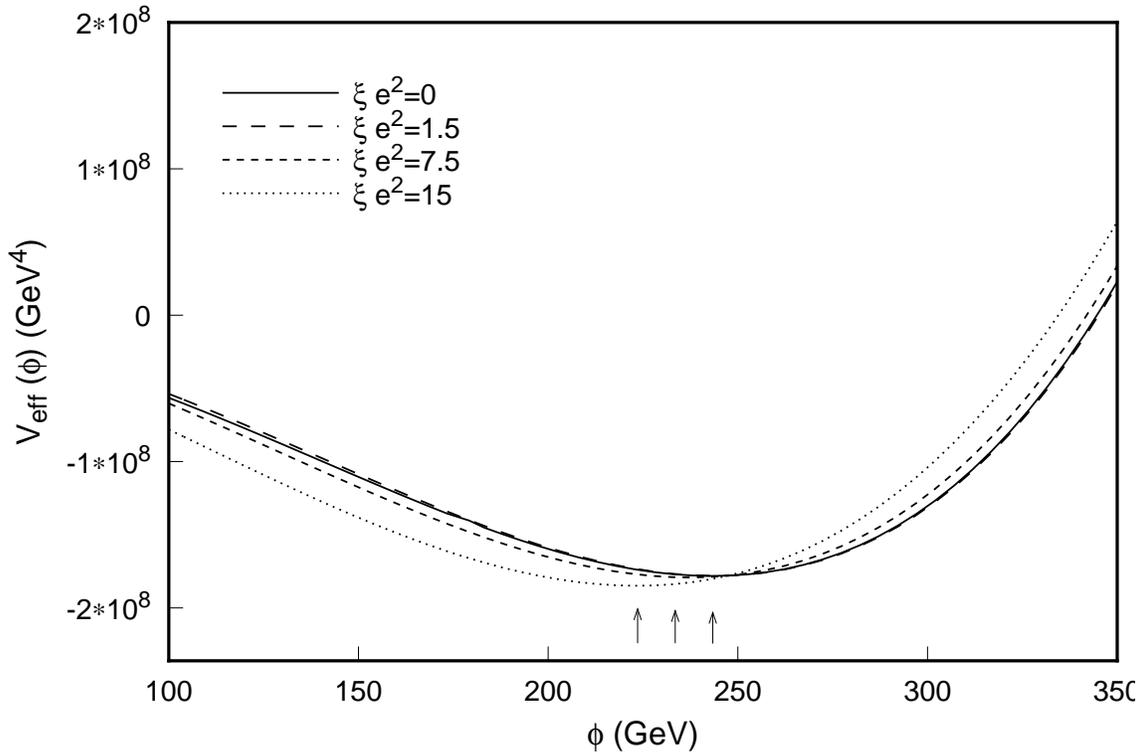} 
\caption{Effective potentials in the $\rxi$ gauge for several values of the gauge parameter.  The arrows indicate the locations of the minima. }
\label{variousrxi} 
\end{figure}
Figures \ref{variousrxi} and \ref{variousxi_V1} illustrate the gauge dependence of the elementary
field effective potential (prior to RG improvement) for $\lambda_i=0.2$. 
Fig. \ref{variousrxi} shows the effective potential in the $\rxi$ gauge for 
several values of the gauge parameter and explicitly indicates the shift in the location of the `electroweak' minimum as a function of the gauge parameter $\xi e^2$.  
In Fig. \ref{variousxi_V1} the one-loop corrections are isolated to highlight the gauge dependence.  
In Fig. \ref{unimprovedvariousa} we plot the one-loop composite operator effective potential with ${\bf p}$-
space smearing function $\tilde{K}(\vec{p})=\theta(\rho-|\vec{p}|)$ for several values of $\rho$.  Since the mass
parameter $m$, which introduces a typical energy scale for the theory, has been taken to be $110 \,{\rm GeV}$,
we expect that for $\rho$ much smaller than this the composite operator will be insensitive to all but 
SSB effects, which are purely infrared.  For $\rho$ much larger than this the composite operator will
detect the shorter-wavelength fluctuations not associated with SSB.  As is shown in the figure, curves with
$\rho$ of order the mass parameter are very close to
 the $\rho=0$ curve.  Only for $\rho \sim 1000 \,{\rm GeV}$ does the  shape
of the composite operator effective potential deviate significantly from the $\rho=0$ curve, so we 
are confident in the ability of the composite operator to provide a reliable separation of the physical scales of
the problem and to isolate only the SSB effects. 
Fig. \ref{lambdai_vs_sVI} shows the results of a vacuum stability bound calculation using the composite 
operator effective potential with $\rho=0$.  For a choice of $s_{VI}$ apply the boundary condition $\lef(s_{VI})=\lambda(s_{VI}) + \Delta\lambda=0$ and use the RG equation to run $\lambda(s)$ down
to $s=0 (\chi=\chi_i)$ (we neglect the mass parameter in our expression for $\Delta\lambda$). The resulting $\lambda_i$ is the minimum value for which the RG-improved effective
potential remains positive up to scale $s_{VI}$.  In a complete treatment this 
lower bound on $\lambda_i$ could be converted to a lower bound on the Higgs pole mass.  We have performed
a similar calculation for the elementary field effective potential in Landau gauge and find the results virtually 
indistinguishable.  For the case studied here, with weak
 gauge couplings, this is simply a reflection of the small quantitative contribution
 from the gauge sector, and not a statement of identity of Landau and Coulomb effective
 potentials. 
 This  suggests
that in the Standard Model as well the Landau gauge results may be similar to those obtained using a gauge-invariant method.  
However, due to numerical differences in the $\beta$ functions, the effect of QCD in the running of $g_y^2$, and of course
the different gauge group of the Standard Model, no immediate conclusions can be drawn in that context.  
In Fig. \ref{Landau_Coulomb_V1} we plot the one-loop correction to the elementary-field effective potential in
Landau gauge and the composite operator effective potential with $(\rho=0)$.  The fact that the curves
are so close in the region in which we choose initial scale $\chi_i$ explains why the corresponding $\Delta\lambda$ and thus the $\lambda_{i,{\rm min}}$ curves are so similar for the two effective potentials.

\section{Composite Operator Effective Potentials for Nonabelian Gauge Theories}

    In an abelian gauge theory, the use of Coulomb gauge was facilitated by
 the observation that a smeared string corresponding to a dipole field
 (cf. (25-27)) yields a gauge-invariantizing factor for a bilocal operator
 which automatically reduces to unity in Coulomb gauge. The need for
 path ordering of such factors in the nonabelian case means that this procedure
 fails and we are forced to seek a more convenient physical gauge. At first sight,
 axial gauge would seem to fit the bill- in this gauge, there is a positive definite
 state space and a well defined Hamiltonian, and the ordinary straight line
 string factor can be used to define a gauge-invariant composite field in the
 usual way
\bdm 
\chi_{1D}\left( t, x \right)=\int{ dl \> \phi^*\left( t, \vec{x}+ {l \over 2} \hat{z} \right)K\left( l \right) {\cal P} 
e^{i \int{dl^\prime \vec{A}(l^{\prime}) \cdot \hat{z} }} \phi^*\left( t, \vec{x}-  {l \over 2} \hat{z} \right)}
\edm
Here the scalar fields have been fixed to lie along an arbitrarily chosen line $\hat{z}$.  
The path ordering along the one-dimensional (spacelike) path is unambiguous.  If the calculation
is then performed in the $A \cdot \hat{z}=0$ spacelike axial gauge the string operator again reduces to unity.

  The use of axial gauge in this fashion leads however to a number of interesting
 delicate points, which divide into two categories- perturbative ultraviolet problems in defining
 the effective potential, and nonperturbative infrared problems associated with the loss of
 long range order \cite{Frohlich}
 when such string operators are employed as order parameters in a 
 broken gauge theory. 

    First, we discuss the problems arising with the definition of a finite composite potential
 in perturbation theory. These problems are not specific to gauge theory - arising simply
 from the UV singularities of one-dimensionally smeared bilocal operators- and can be
 illustrated in the simple scalar model of Section 2.  We found there that the one loop 
 composite effective potential yields a term 
\bdm
\int{\frac{d^{3}p}{(2\pi)^{3}}(\sqrt{\vec{p}^{2}+2\lambda_{R}\chi+
(m_{R}^{2}+\lambda_{R}\chi)(1-\tilde{K}(\vec{p}))}-|\vec{p}|)}
\edm
  where $\tilde{K}(\vec{p})$ is the Fourier transform of the coordinate space
 bilocal smearing function $K(\vec{r})$. It is crucial that the terms proportional to
 powers of $\tilde{K}$ in the expansion of this expression in powers of $\chi$ 
 {\em not} introduce additional ultraviolet divergences, as there are no counterterms
 available in the Lagrangian to absorb them. If we simply point split the fields (along
 the z direction, say)  by
 taking $K(\vec{r})=\frac{1}{2}\delta(x)\delta(y)(\delta(x-1/\rho)+\delta(x+1/\rho))$,
 corresponding to $\tilde{K}=\cos{\frac{p_z}{\rho}}$, we find that the one-loop
 potential expanded in $\chi$ is UV finite at  ${\cal O}(\chi)$ but logarithmically divergent
 at  ${\cal O}(\chi^2)$. Namely:
\begin{eqnarray}
\label{eq:axdiv1}
  \int d^{3}p\frac{\cos(p_z/\rho)}{\sqrt{\vec{p}^{2}+m_{R}^{2}+3\lambda_{R}\chi}}&<&\infty \\
  \int d^{3}p\frac{\cos^{2}(p_z/\rho)}{(\vec{p}^{2}+m_{R}^{2}+3\lambda_{R}\chi)^{3/2}}&\sim&
 \frac{1}{2}\int d^{3}p\frac{1}{(\vec{p}^{2}+m_{R}^{2}+3\lambda_{R}\chi)^{3/2}}+{\rm finite}  \\
&\sim&{\rm log\;\;divergent}
\end{eqnarray}
 The origin of the divergence is not hard to find- at order $\chi^{2}$, the momentum space
 double  insertion of the point split operator  $\phi(\vec{r}+\frac{1}{\rho}\hat{z})\phi(\vec{r}-
\frac{1}{\rho}\hat{z})$ leads to a UV divergence when the coordinate space integration
 defining the Fourier transform brings both pairs of field operators simultaneously together.
 Clearly, holding the two fields at a fixed separation is a prescription for trouble. On the
 other hand, taking a smooth smearing function such as $\tilde{K}=e^{-p_{z}^{2}/4\rho^{2}}$
 (in analogy to the 3 dimensional Gaussian smearing of Sections 2,3), leads to an
uncompensated divergence at  O($\chi$), with order $\chi^{2}$ and higher terms finite.
 The linear divergence in $\chi\int d^{3}p\frac{e^{-p_{z}^{2}/4\rho^{2}}}{\sqrt{\vec{p}^{2}+m_{R}^{2}+3\lambda_{R}\chi}}$
 is also readily understood in operator terms. Here the one-dimensional smearing integral
 leads  directly to a linear divergence due to the quadratic short-distance divergence of
 a product of two scalar field operators approaching the same point.  

   The solution to these ultraviolet problems is actually quite simple- one need only choose
 a smearing function which vanishes at the origin in coordinate space. This removes the
 divergence from the one dimensional integral in the region where the two field points
 approach one another in a single insertion of the composite field $\chi$, while removing
 the logarithmic divergence at order $\chi^{2}$ by smearing out the split field locations.
 For example, a suitably normalized smearing function is
\begin{equation}
\label{eq:axsmear}
 K(\vec{r})=\frac{2}{\sqrt{\pi}}\rho^{3}\delta(x)\delta(y)z^{2}e^{-\rho^{2}z^{2}}
\end{equation}
 As for the three dimensional smearings of Sections 2-4, one should take the 
 smearing parameter $\rho$ small, corresponding to widely separated peaks in
 $K(\vec{r})$. In the opposite limit of $\rho$ large, the UV divergences will reappear,
 distorting the shape of the effective potential. Of course, the requirement that
 $K(0)=0$ now means that the Fourier transform $\tilde{K}$ is not positive for
 all $\vec{p}$, so there is no rigorous argument that the natural domain of $\chi$
 is restricted to the positive real axis. Still, for small $\rho$ in the symmetry-broken
 phase, we expect that the minimum of the effective potential will be found at a safely 
 positive location, so this property is perhaps not terribly important.

   A further difficulty which one encounters in employing $\chi_{1D}$ as an 
 order parameter is specific to gauge theories with nontrivial topological
 structure. The decorrelating effects induced by instantons on bilocal operators
 with a gauge-invariantizing string was first pointed out by Fr\"ohlich et. al. \cite{Frohlich}.
  The existence of Gribov copies \cite{Nill} can also lead to a destruction of long range order,
 again as a result of large nonperturbative field configurations. All such effects
 are presumably of order $e^{-c/\hbar}$ and therefore not visible in a standard 
 perturbative loop expansion. From a practical point of view, in the 
 weakly coupled electroweak case
 instanton effects  (unenhanced by large combinatoric prefactors as is 
 potentially the case in  multiparticle
 production at high energy) are extremely small, so these effects are clearly ignorable.
 In strongly coupled theories- when studying dynamical symmetry breaking, for
 example- the problem returns and caution will be required when taking the limit
 $\rho\rightarrow 0$ in an axial type gauge. We should remind the reader that for the 
 Coulomb gauge smeared bilocal operator, Kennedy and King have shown \cite{KK}
 that this limit  leads to a nonvanishing order parameter in the symmetry broken phase. 

\section{Conclusions}
The computation of lower bounds on the Higgs mass from vacuum stability constraints using the elementary field
effective potential results results in  unphysical gauge dependence in such bounds.  
We have formulated
an effective potential in terms of  gauge invariant composite operators which avoids the ultraviolet problems of 
local composite operators while retaining an energy interpretation.  
We have shown how this could be used to calculate
the vacuum stability bound on the Higgs mass in the context of a toy model, the abelian Higgs model coupled
to a fermion, and for the set of parameters chosen (corresponding to a weakly coupled
 gauge sector) find that the results are quantitatively close to those obtained 
from the elementary field effective potential in Landau gauge.  
While the extension to a nonabelian gauge theory using a related composite operator in axial gauge 
introduces additional subtleties, we see no serious obstacles to the complete calculation.
Thus, the tool might be extended to models of phenomenological interest, such as the Standard Model
or its extensions. Composite effective potentials are essential in the study of
  dynamical symmetry breaking, so the issues discussed in this paper should prove of
 value in that context as well.
\vspace{1.5in}
\section{Acknowledgements}
  The authors are grateful for a useful communication from E. Seiler, and for 
 conversations with D. Boyanovsky, Paresh Malde and E. Weinberg.
 This work was supported in part by the  NSF through grant 93-22114. 

\newpage
\section*{Appendix A:  $\msb$ renormalization}

\subsection*{$\msb$ counterterms}
\bea
\delta Z_{\phi}&=&\left( -2 e^2 + 2 g_y^2\right)\pole  \nonumber \\
\delta Z_{\lambda}&=&\left( -10 \lambda + 6 e^2 - 3 {{e^4} \over {\lambda}}
- 4 g_y^2 + 4 {{g_y^4} \over {\lambda}} \right)\pole  \nonumber \\
\delta Z_{g_y}&=&\left({3 \over 4} e^2 - 2 g_y^2 \right)\pole \nonumber \\
\delta Z_{m^2}&=&\left(  - 4 \lambda + 3 e^2 - 2 g_y^2\right)\pole \nonumber \\
\delta Z_e&=&    -{1 \over 3} e^2 \pole \nonumber\\
\delta Z_{A}&=&  {2 \over 3} e^2 \pole \nonumber\\
\delta Z_t&=& g_y^2 \pole \nonumber \\
\eea

\subsection*{$\beta$ functions}

The relevant one-loop $\msb$ $\beta$ and $\gamma$ functions for the theory are:
\bea
{\beta}_{\lambda} &=&{1 \over {16 \pi^2}} \left( 20 \lambda^2 + 6 e^4 - 8 g_y^4 - 12 e^2 \lambda + 8 \lambda g_y^2 \right)\\
{\beta}_{e} &=& {1 \over {16 \pi^2}}  \left( {2 \over 3} e^3 \right)\\
{\beta}_{g_y} &=& {1 \over {16 \pi^2}} g_y \left(- {3 \over 2} e^2 + 4 g_y^2 \right) \\
{\gamma}_{\phi} &=& {1 \over {16 \pi^2}} \left(-e^2 + g_y^2  \right)
\eea

\end{document}